\pdfoutput=1

\documentclass[aps,twocolumn,amsmath,amssymb,preprintnumbers,floatfix,prl,longbibliography]{revtex4-2}

\usepackage[utf8]{inputenc}
\usepackage{newtxtext}
\usepackage[upint]{newtxmath}
\usepackage{microtype}
\usepackage{bm}

\usepackage{enumerate}
\usepackage{amsfonts}
\usepackage{amsmath}
\usepackage{amssymb}
\usepackage{siunitx}
\usepackage{braket}
\usepackage{color}
\usepackage{soul}

\usepackage{graphicx}

\usepackage[colorlinks,allcolors=blue]{hyperref}
\usepackage[capitalize]{cleveref}

\usepackage{lipsum}
\usepackage{comment}
\usepackage{todonotes}
\setuptodonotes{inline}

\definecolor{DarkRed}{rgb}{0.65,0,0}%
\definecolor{Green}{rgb}{0,0.3,0.3}
\definecolor{Purple}{rgb}{0.3,0,0.65}
\definecolor{Red}{rgb}{1,0,0}
\definecolor{Blue}{rgb}{0,0,0.85}
\definecolor{Magenta}{rgb}{1,0,1}

\newcommand{\ve}[1]{\boldsymbol{#1}}

\newcommand{\vk}{{\ve{k}}} 

\newcommand{\be}{\begin{equation}}
		\newcommand{\ee}{\end{equation}}

\newcommand{\dn}{\downarrow}
\newcommand{\up}{\uparrow}
\newcommand{\ph}{{\phantom{\dagger}}}

\newcommand{\prlsection}[1]{\textit{#1}.\kern0.05em---\kern0.05em\ignorespaces}

\begin{document}
\title{dc Josephson Effect in Altermagnets}
\author{Jabir Ali Ouassou, Arne Brataas, and Jacob Linder}
\affiliation{Center for Quantum Spintronics, Department of Physics, Norwegian \\ University of Science and Technology, NO-7491 Trondheim, Norway}

\begin{abstract}
	The ability of magnetic materials to modify superconducting systems is an active research area for possible applications in thermoelectricity, quantum sensing, and spintronics.
	We consider the fundamental properties of the Josephson effect in a third class of magnetic materials beyond ferromagnets and antiferromagnets: altermagnets.
	We show that despite having no net magnetization, altermagnets induce 0-$\pi$ oscillations.
	The decay length and oscillation period of the Josephson coupling are qualitatively different from ferromagnetic junctions and depend on the crystallographic orientation of the altermagnet.
	The Josephson effect in altermagnets thus serves a dual purpose: it acts as a signature that distinguishes altermagnetism from conventional (anti)ferromagnetism and offers a way to tune the supercurrent via flow direction anisotropy.
\end{abstract}
\maketitle

\begin{figure}[t]
	\includegraphics[width=0.87\columnwidth]{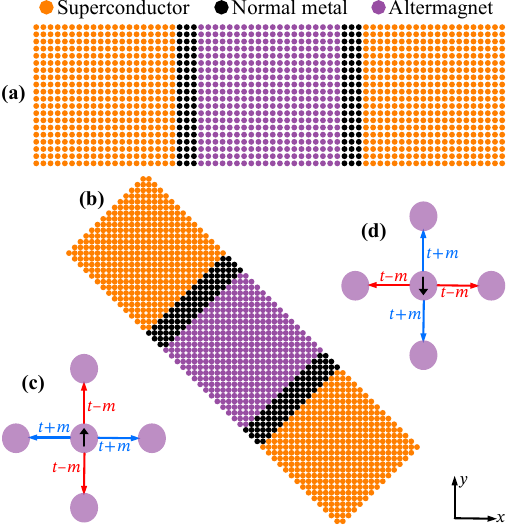}
	\caption{
		(a--b)~Illustration of the Josephson junctions considered here.
		Each circle represents one atom in a 2D square lattice in real space.
		In an experiment, there is likely a lattice mismatch between the different regions at the interfaces, which can reduce the supercurrent amplitude.
		(a)~``Straight junctions'' are aligned with the crystallographic axes,
		and thus have the interface normals $\bm{n} = \bm{e}_x$.
		(b)~``Diagonal junctions'' have 45$^\circ$ misalignment relative to the lattice, so $\bm{n} = (\bm{e}_x - \bm{e}_y)/\sqrt{2}$.
		(c--d)~Illustration of the altermagnetic order parameter~$\bm{m}_{ij}$.
		(c)~Spin-up electrons have increased hopping amplitudes $t \rightarrow t + m$ along the $x$ axis,
		and decreased hopping amplitudes $t \rightarrow t - m$ along the $y$ axis.
		(d)~For spin-down electrons, the situation is exactly reversed.
	}
	\label{fig:model}
\end{figure}

\textit{Introduction.}
Spin splitting of quasiparticle bands in condensed matter systems is a crucial functional property of materials explored in spintronics \cite{hirohata_jmmm_20}.
Recent works detail mechanisms for such splitting distinct from ferromagnetic and relativistically spin-orbit coupled systems \cite{yuan_prb_20, smejkal_prx_22}, originally envisioned in Ref.~\cite{pekar_zetf_64}.
One such mechanism consists of a spin-lattice coupling due to an internal periodic magnetic field in the material \cite{yuan_prb_20}, which ultimately leads to a sizeable momentum-dependent spin splitting in the material.
It is important to note that such splitting occurs even when disregarding conventional atomic spin-orbit coupling, which has a relativistic origin, the latter thus being much weaker than the new mechanism considered in recent works.
The materials displaying this type of magnetic properties are known as altermagnets.
They have a large $\vk$-dependent spin splitting of the bands, which is even in powers of $\vk$, and that persists in the absence of relativistic spin-orbit couplings \cite{smejkal_prx_22}.
\textit{Ab initio} calculations have identified several possible material candidates that can host an altermagnetic state, including metals like RuO$_2$ and Mn$_5$Si$_3$ \cite{smejkal_sciadv_20, reichlova_arxiv_20, smejkal_prx_22} as well as semiconductors/insulators like MnF$_2$ and La$_2$CuO$_4$ \cite{yuan_prb_20, lopez-moreno_pccp_16}.

Through the proximity effect, metallic materials which are not superconducting can inherit the two fundamental characteristics of superconductors: the Meissner effect \cite{clarke_prsa_69} and dissipationless charge flow \cite{oda_ssc_80}.
When magnetic materials become superconducting via the proximity effect, both the Meissner effect and the dissipationless transport change in qualitatively new ways.
For instance, the Josephson coupling through ferromagnetic materials displays an effect known as 0-$\pi$ oscillations \cite{ryazanov_prl_01, buzdin_rmp_05}.
This means that the ground-state phase difference between the superconductors alternates between 0 and $\pi$, depending on the junction parameters.
As a result, the supercurrent vanishes at certain lengths and temperatures.
Such $\pi$ junctions can be used for qubits \cite{ioffe_nature_99, feofanov_natphys_10}, and have also been generalized to $\phi_0$ junctions \cite{geshkenbein_pisma_86, millis_prb_88, szombati_natphys_16} where the system acts as a quantum phase battery supplying an arbitrary phase between 0 and $\pi$.

In this work, we consider Josephson junctions with altermagnetic interlayers (see \cref{fig:model}).
Surprisingly, we find that despite the absence of any magnetization in altermagnets, the supercurrent displays 0-$\pi$ oscillations.
The behavior is different from both ferromagnetic and antiferromagnetic Josephson junctions.
In the latter scenario, the $\pi$-state occurs in the very specific case of a junction with exactly an odd number of atoms~\cite{andersen_prl_06}, so that a net magnetic moment exists.
In addition, we find that both the decay and oscillation period of the supercurrent in the altermagnetic case exhibits anisotropy with respect to the crystallographic orientation of the interface relative the superconductors.
These unique characteristics of the Josephson current in altermagnets can be used as a tool to identify the altermagnetic state among the list of candidate materials that have recently been identified through \textit{ab initio} calculations \cite{smejkal_prx_22}.

\textit{Model.}
As shown in \cref{fig:model}, we consider two kinds of Josephson junctions.
Both are created from a 2D square lattice with lattice constant~$a$, but have different junction orientations relative to the crystallographic axes.
At the ends of each junction is a $20a\times20a$ BCS superconductor.
The two superconductors that form each Josephson junction have a variable phase difference~$\delta\!\varphi$.
Next to the superconductors are {thin} normal-metal spacers of lengths $3a$ (straight junctions) or $3\sqrt{2}a$ (diagonal junctions).
Finally, the center of each junction is an altermagnet of varying length $L \in [0, 40a]$.
In \cref{fig:model}, we plot an intermediate junction length $L = 20a$.

To model the proposed physical setup we employ the Bogoliubov--de\,Gennes (BdG) method \cite{zhu2016, degennes1966}.
Our starting point is a mean-field tight-binding Hamiltonian that includes altermagnetism and conventional superconductivity:
\begin{equation}
	\begin{aligned}
		\mathcal{H}
		 & = E_0
		- \sum_{i\sigma} \mu_i c^\dag_{i\sigma} c^\ph_{i\sigma}
		- \sum_i (\Delta_i^{\vphantom{*}} c^\dag_{i\dn} c^\dag_{i\up} + \Delta_i^* c_{i\up} c_{i\dn}) \\
		 & \phantom{==}
		- \sum_{\langle i, j \rangle \sigma} t_{ij} c^\dag_{i\sigma} c^\ph_{j\sigma}
		- \sum_{\langle i, j \rangle \sigma\sigma'} (\bm{m}_{ij} \cdot \bm{\sigma})_{\sigma\sigma'} c^\dag_{i\sigma} c^\ph_{j\sigma'},
	\end{aligned}
\end{equation}
where $c^\ph_{i\sigma}$ and $c^\dag_{i\sigma}$ are the usual electronic annihilation and creation operators, and $\bm{\sigma} = (\sigma_1, \sigma_2, \sigma_3)$ is the Pauli vector.
$E_0$ describes a constant contribution which is not important for the non-selfconsistent calculations below.
We choose constant nearest-neighbor hopping amplitudes~$t_{ij} \equiv t$ and chemical potentials~$\mu_i = -t/2$.
In the two superconductors, we set $\Delta_i = \Delta e^{\pm i \delta\!\varphi/2}$.
The gap was calculated using the interpolation formula $\Delta(T) \approx \Delta(0) \tanh\big[1.74 \sqrt{T_c/T - 1}\big]$,
where we chose a zero-temperature gap $\Delta(0) = t/10$.
The critical temperature was determined using the BCS ratio $\Delta(0)/T_c \approx 1.764$.
In the altermagnet, we set $\bm{m}_{ij} = +m\bm{e}_z$ for nearest-neighbor hopping along the $x$ axis and $\bm{m}_{ij} = -m\bm{e}_z$ for hopping along the $y$ axis.
This corresponds to a low-energy effective Hamiltonian $mk_xk_y\sigma_z$ or $m(k_x^2-k_y^2)\sigma_z$, depending on the crystallographic orientation of the sample.
This Hamiltonian differs from both the momentum-independent spin-splitting $m\sigma_z$ of a ferromagnet and a Rashba-type spin-orbit coupling
$mk_{x(y)}\sigma_z$.

The model above has three parameters that were varied between simulations:
The altermagnet length $L \in [0, 40a]$, the magnitude of its order parameter $m \in \{0.5\Delta, 1.5\Delta, 0.5t, 0.9t\}$, and the phase difference $\delta\!\varphi \in \{0, 0.02\pi, \ldots\}$.
For each combination of these parameters, we calculated the Josephson supercurrent~$I$ flowing along the junction using the methodology described below.
The current-phase relation $I(\delta\!\varphi)$ for each junction was then fit to Fourier sine series,
\begin{equation}
	I(\delta\!\varphi) = \sum_{n>0} I_n \sin(n\,\delta\!\varphi).
\end{equation}
The amplitude of the first harmonic $I_1$ was extracted from these fits, and used to judge whether the Josephson junction is in a 0-state or $\pi$-state based on its sign.
In general, it is also possible to have $\varphi_0$ junctions, where also cosine terms need to be included in the Fourier expansion; however, no sign of such $\varphi_0$ effects were found in any of our simulations.
While the first harmonic is ideal for locating 0-$\pi$ transitions, we also calculated the critical current $I_c \equiv \text{max}_{\delta\!\varphi}\, |I(\delta\!\varphi)|$ for some interesting junctions as it is more experimentally accessible.

\textit{Methodology.}
The fermionic operators at each site~$i$ can be grouped into Nambu vectors $\hat{c}_i \equiv (c_{i\up}^\ph, c_{i\dn}^\ph, c_{i\up}^\dag, c_{i\dn}^\dag)$, which may in turn be collected into a $4N$-element vector $\check{c} \equiv (\hat{c}_1, \ldots, \hat{c}_N)$ containing every fermionic operator on the lattice.
The Hamiltonian operator can then be expressed via a $4N\times4N$ Hamiltonian matrix: ${\mathcal{H} = E_0 + \frac12 \check{c}^\dag \check{H} \check{c}}$.
The most common approach to solving the BdG equations consists of diagonalizing~$\check{H}$, and then expressing physical observables of interest in terms of its eigenvectors and eigenvalues.
However, an alternative approach has gained momentum over the last decade: The Kernel Polynomial Method \cite{weisse2006, covaci2010, nagai2012}.
Instead of diagonalizing the Hamiltonian, one calculates a Green function matrix from the Hamiltonian matrix, which can be done efficiently and accurately using a series expansion in Chebyshev polynomials.
Many physical observables of interest can then be directly extracted from the elements of this Green function.

\begin{figure*}
	\includegraphics[width=\textwidth]{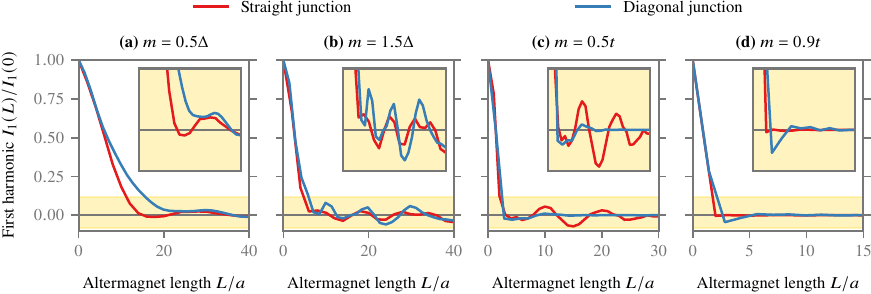}
	\caption{
		First harmonic~$I_1$ of the Josephson supercurrent $I(\delta\!\varphi)$ as function of the altermagnet length $L$ for the junctions in \cref{fig:model}.
		To simplify the comparison between the two different geometries, each curve was normalized to the amplitude $I_1(0)$ in the absence of the altermagnetic interlayer.
		As indicated above the plots, different panels and curves correspond to different altermagnetic order parameters and junction types, respectively.
		The insets zoom in on the regions in the golden boxes, in order to highlight the $0$--$\pi$ oscillations for large junction lengths.
	}
	\label{fig:supercurrent}
\end{figure*}

There are many variants of the Chebyshev methods outlined above.
We here use the Fermi operator expansion method~\cite{goedecker1999}, which for the case of the BdG Hamiltonian is explained in detail in Ref.~\cite{benfenati2022}.
The starting point is then the \emph{Fermi matrix} $\check{F} \equiv f(\check{H})$, where $f(\epsilon) = [1+\exp(\epsilon/T)]^{-1}$ is the Fermi--Dirac distribution at temperature~$T$.
The function~$f$ should be interpreted in terms of its Taylor expansion when applied to the matrix~$\check{H}$.
Using the kernel polynomial method, we can expand the Fermi matrix in Chebyshev polynomials as
\begin{equation}
	\check{F} = \frac{1}{2} f_0 \check{I} + \sum_{m=1}^{M-1} f_m g_m \check{T}_m,
\end{equation}
where $f_m$ are the Chebyshev moments of the Fermi--Dirac distribution~\cite{benfenati2022, weisse2006}, $g_m$ are the Jackson kernel coefficients~\cite{weisse2006}, $\check{T}_m \equiv T_m(\check{H})$ are Chebyshev matrix polynomials~\cite{weisse2006}, and $\check{I}$ is the identity matrix.
Note that the above assumes that $\check{H}$ has been normalized such that all its eigenvalues have magnitudes below unity; this is in practice easily achieved by scaling $\check{H}$ by its 1-norm.
The Chebyshev polynomials are calculated via the usual recursion relation ${\check{T}_0 = \check{I},\, \check{T}_1 = \check{H},\, \check{T}_m = 2\check{H} \check{T}_{m-1} - \check{T}_{m-2}}$~\cite{weisse2006}.
Calculating $\{\, \check{T}_m \}$ is the computationally limiting part of our calculation, but was significantly sped up using sparse matrices with fully-parallelized block-wise matrix multiplication.
All simulations presented here were performed using $M = 4000$ Chebyshev moments.
We found that this provides negligible truncation error for typical junctions, which is consistent with the findings for LDOS calculations in e.g.\ Ref.~\cite{covaci2010}.
The calculations were performed at a temperature $T = T_c/20$.

The procedure above provides us with a $4N\times4N$ Fermi matrix~$\check{F}$.
This can be deconstructed into $4\times4$ blocks in Nambu space, $\check{F} = [\hat{F}_{ij}]$.
Following a similar approach as Ref.~\cite{benfenati2022}, it can then be shown that the elements of these matrices are:
\begin{equation}
	\hat{F}_{ij} =
	\begin{pmatrix}
		\braket{c_{j\uparrow}^{\dag} c_{i\uparrow}^\ph}      & \braket{c_{j\downarrow}^{\dag} c_{i\uparrow}^\ph}      & \braket{c_{j\uparrow}^\ph c_{i\uparrow}^\ph}      & \braket{c_{j\downarrow}^\ph c_{i\uparrow}^\ph}      \\[1ex]
		\braket{c_{j\uparrow}^{\dag} c_{i\downarrow}^\ph}    & \braket{c_{j\downarrow}^{\dag} c_{i\downarrow}^\ph}    & \braket{c_{j\uparrow}^\ph c_{i\downarrow}^\ph}    & \braket{c_{j\downarrow}^\ph c_{i\downarrow}^\ph}    \\[1ex]
		\braket{c_{j\uparrow}^{\dag} c_{i\uparrow}^{\dag}}   & \braket{c_{j\downarrow}^{\dag} c_{i\uparrow}^{\dag}}   & \braket{c_{j\uparrow}^\ph c_{i\uparrow}^{\dag}}   & \braket{c_{j\downarrow}^\ph c_{i\uparrow}^{\dag}}   \\[1ex]
		\braket{c_{j\uparrow}^{\dag} c_{i\downarrow}^{\dag}} & \braket{c_{j\downarrow}^{\dag} c_{i\downarrow}^{\dag}} & \braket{c_{j\uparrow}^\ph c_{i\downarrow}^{\dag}} & \braket{c_{j\downarrow}^\ph c_{i\downarrow}^{\dag}} \\
	\end{pmatrix}.
	\label{eq:fermi-matrix-comp}
\end{equation}
This implies that any physical observable which can be calculated from two-point finite-temperature correlation functions on the lattice can be calculated directly from the Fermi matrix.

We evaluate the charge current inside the normal metal spacer.
Charge conservation ensures that the current is constant anywhere along the junction in a stationary system.
We compute the charge current by summing over bond currents.
The bond current between two sites $i$ and $j$ can be written~\cite{zhu2016}
\begin{equation}
	J_{ij} = ie \sum_{\sigma} \left( t_{ij} \braket{c^\dag_{i\sigma} c^\ph_{j\sigma}} - t_{ji}\, \braket{c^\dag_{j\sigma} c^\ph_{i\sigma}} \right),
\end{equation}
where $e < 0$ is the electron charge.
By comparison with \cref{eq:fermi-matrix-comp}, we see that the bond current $J_{ij}$ can be trivially calculated from appropriate traces of $\hat{F}_{ij}$ and $\hat{F}_{ji}$.
The bond current along the junction direction~$\bm{n}$ is then simply $J_{ij}(\bm{\delta}_{ij}\cdot\bm{n})$, where $\bm{\delta}_{ij}$ is a unit vector that points from site~$i$ towards site~$j$.
The total current~$I$ flowing through the junction is found by integrating this over a cross section of the junction.

\textit{Results and discussion.}
The main results of our numerical simulations are in \cref{fig:supercurrent}.
First, we observe that 0-$\pi$ oscillations are possible in \emph{both} straight and diagonal junctions.
This finding is interesting since such oscillations are typically found in Josephson junctions with \emph{magnetic} interlayers, whereas altermagnets have zero magnetization.
Moreover, 0-$\pi$ oscillations do not appear in Rashba spin-orbit coupled junctions either, which have a different spin-momentum coupling (odd-in-momentum) compared to altermagnets.
Second, we see that the 0-$\pi$ oscillations behave qualitatively differently from ferromagnetic Josephson junctions: the latter typically has an exponential decay with superimposed oscillations, whereas in the altermagnet case there is an initial large decay followed by oscillations with a much weaker damping.
This result is most striking in \cref{fig:supercurrent}(b), where we find a pure decay at $L < 8a$ followed by nearly pure oscillation at $L > 10a$.

Physically, the oscillations in straight junctions can be understood as follows.
Conventional superconductivity consists of singlet Cooper pairs $\ket{\up\dn} - \ket{\dn\up}$.
As the Cooper pairs leak into the altermagnet along the $x$ axis, spin-up electrons have a hopping amplitude $t+m$ while spin-down electrons have a hopping amplitude $t-m$ (see \cref{fig:model}).
This ``speed difference'' causes position-dependent phase differences between spin-up and spin-down electrons.
Such spin-dependent phase shifts are well-known to cause 0-$\pi$ oscillations from previous studies on ferromagnetic Josephson junctions.
For diagonal junctions, however, the most direct path between the superconductors consists of an equal number of hops along the $x$ and $y$ axes.
Since the electrons experience opposite spin-dependent phase shifts in these two cases, their effects appear to partially cancel.

\Cref{fig:supercurrent} shows that the initial decay in $I_1(L)$ is in general accelerated as $m$ is increased.
However, 0-$\pi$ oscillations are found over a much wider parameter range for straight than diagonal junctions, consistent with the discussion above.
For example, for small altermagnetic order parameters $m = 0.05t$ [\cref{fig:supercurrent}(a)], the first 0-$\pi$ oscillation occurs already at $L = 15a$ for the straight junction, but not until $L = 35a$ for the diagonal junction.
On the other hand, for large altermagnetic order parameters $m = 0.5t$ [\cref{fig:supercurrent}(c)], the first 0-$\pi$ oscillation occurs simultaneously, but sustained 0-$\pi$ oscillations for a large range of junction lengths is found only for straight junctions.
It is only for intermediate values $m = 0.15t$ [\cref{fig:supercurrent}(b)] that we find qualitatively similar results in straight and diagonal junctions.
Finally, for very large values $m = 0.9t$, the supercurrent decays extremely fast for both straight and diagonal junctions, limiting the number of visible oscillations for both junction types.

For very large altermagnetic order parameters $m \rightarrow t$, spin-down electrons become nearly immobile along the $x$ axis.
In this limit, spin-zero Cooper pairs clearly cannot propagate through the altermagnet, and the Josephson effect vanishes.
This explains the extremely sharp decay in \cref{fig:supercurrent}(d).
Interestingly, if an altermagnet with $m \rightarrow t$ could be realized experimentally, this might also serve as a new kind of filter for spin-triplet Cooper pairs.
Specifically, we would expect $\ket{\up\up}$ pairs to only move along the $x$ axis, $\ket{\dn\dn}$ pairs to only move along the $y$ axis, and any $\ket{\up\dn}\mp\ket{\dn\up}$ pairs to decay.
This may thus provide a non-destructive way to separate the different equal-spin-triplet pairs generated in superconducting spintronics while eliminating any remaining spin-zero pairs.

\begin{figure}[t]
	\includegraphics[width=0.85\columnwidth]{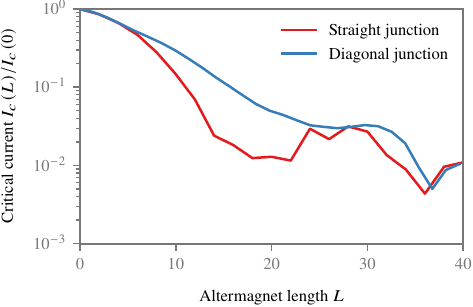}
	\caption{
		Critical current~$I_c$ as a function of the altermagnet length~$L$ for $m = 0.5\Delta$ and $T = 0.05T_c$ [cf. \cref{fig:supercurrent}(a)].
	}
	\label{fig:critcurrent-L}
\end{figure}

In \cref{fig:critcurrent-L}, we show the critical current, {rather than just the first harmonic,} for one of the junctions considered {$(m=0.5\Delta)$}.
In \cref{fig:supercurrent}(a) we saw {that this choice of $m$} produces 0-$\pi$ oscillations at $L=15a$ and $L=21a$ for straight but not diagonal junctions, which causes significant $I_c$ suppression.
However, due to the presence of higher harmonics, it is difficult from the $I_c$ curve alone {in \cref{fig:critcurrent-L}} to {observe} that there are two such 0-$\pi$ oscillations in this area.
This becomes even more challenging for higher values of~$m$ (not shown), where qualitatively similar oscillations are observed, but the shorter oscillation period makes it more difficult to determine the exact number of zero crossings.
We also see that both junctions display a 0-$\pi$ oscillation for $L \approx 35a$, which for the diagonal junction is the first 0-$\pi$ oscillation.

\begin{figure}[t]
	\includegraphics[width=0.85\columnwidth]{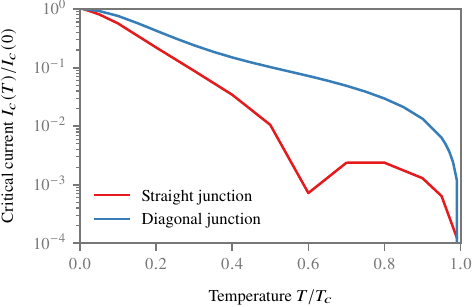}
	\caption{
		Critical current~$I_c$ as a function of temperature~$T$ for $m = 1.5\Delta$ and $L = 12a$ [cf. \cref{fig:supercurrent}(b)].
	}
	\label{fig:critcurrent-T}
\end{figure}

\Cref{fig:critcurrent-T} shows the critical current~$I_c$ vs. temperature~$T$, which is one experimental signature of 0-$\pi$ transitions in Josephson junctions~\cite{ryazanov_prl_01}.
For these calculations, we picked an altermagnet length $L = 12a$ which according to \cref{fig:supercurrent}(b) is close to a 0-$\pi$ transition for straight but not diagonal junctions.
For the straight junction, we find a non-monotonic critical current that dips sharply at $T = 0.6T_c$.
For both $T = 0.5T_c$ and $T = 0.7T_c$, the current-phase relation is completely dominated by the first harmonic~$I_1$.
However, it changes sign between these two points, so the dip is a signature of a 0-$\pi$ transition as a function of temperature.
This is in contrast to the diagonal junction, where we find no 0-$\pi$ transition for these parameters.

\textit{Conclusion.}
In summary, we have demonstrated that the Josephson effect through altermagnets exhibits different properties than in the two conventional classes of magnetic materials, ferromagnets and antiferromagnets.
Despite the absence of a net magnetization, altermagnets induce 0-$\pi$ oscillations in the Josephson effect.
We have also shown that the decay and oscillation period of the Josephson current strongly depend on the crystallographic orientation of the altermagnet relative the superconductors.
The Josephson effect can therefore be used both to distinguish the altermagnet from conventional (anti)ferromagnetism and additionally offers a way to tune the supercurrent via flow direction anisotropy.

\begin{acknowledgments}
	\textit{Acknowledgments.}
	This work was supported by the Research Council of Norway through Grant No. 323766 and its Centres of Excellence funding scheme Grant No. 262633 ``QuSpin.''
	The simulations were performed on resources provided by Sigma2---the National Infrastructure for High Performance Computing and Data Storage in Norway.
\end{acknowledgments}

\end{document}